\documentclass{aa} 
\usepackage{graphics}
\usepackage{amsfonts}

\newcommand{\diff}{\mathrm d}

\newcommand{\eqref}[1]{(\ref{#1})}
\renewcommand{\O}{\mathcal O}

\begin{document}

   \thesaurus{02(12.07.1;  
                 03.13.1;  
                 03.13.4)} 

   \title{A fast direct method of mass reconstruction for
   gravitational lenses}
   \titlerunning{A direct method of mass reconstruction}
   \author{M. Lombardi\inst{1,2} \and G. Bertin\inst{2}}
   \authorrunning{M. Lombardi \& G. Bertin}
   \offprints{M. Lombardi}
   \mail{lombardi@sns.it}
   \institute{European Southern Observatory,
              Karl-Schwarzschild Stra\ss e 2, 
              D 85748 Garching bei M\"unchen, Germany
              \and
              Scuola Normale Superiore,
              Piazza dei Cavalieri 7, I 56126 Pisa, Italy}
   \date{Received 25 March 1999; accepted 17 May 1999}

   \maketitle

   \begin{abstract}
     
     Statistical analyses of observed galaxy distortions are often
     used to reconstruct the mass distribution of an intervening
     cluster responsible for gravitational lensing. In current
     projects, distortions of thousands of source galaxies have to be
     handled efficiently; much larger data bases and more massive
     investigations are envisaged for new major observational
     initiatives. In this article we present an efficient mass
     reconstruction procedure, a direct method that solves a
     variational principle noted in an earlier paper, which, for
     rectangular fields, turns out to reduce the relevant execution
     time by a factor from $100$ to $1000$ with respect to the fastest
     methods currently used, so that for grid numbers $N = 400$ the
     required CPU time on a good workstation can be kept within the
     order of 1 second. The acquired speed also opens the way to some
     long-term projects based on simulated observations (addressing
     statistical or cosmological questions) that would be, at present,
     practically not viable for intrinsically slow reconstruction
     methods.


   \end{abstract}

%

\section{Introduction}

In the context of weak or statistical lensing, the problem of the
determination of the dimensionless mass density distribution
$\kappa(\vec\theta)$ from a map of the reduced shear $g(\vec\theta)$
has been considered in detail by various authors, using either
simulations or analytical calculations (e.g., Bartelmann 1995,
Schneider 1995, Seitz \& Schneider 1996, Squires \& Kaiser 1996,
Lombardi \& Bertin 1998a, 1998b).

The mass inversion is usually performed starting from the vector field
$\vec{\tilde u}(\vec\theta)$ defined in terms of the
\textit{measured\/} reduced shear $g(\vec\theta)$ (Kaiser 1995). In
the ideal case where the \textit{measured\/} shear $g(\vec\theta)$ is
just the \textit{true\/} shear $g_0(\vec\theta)$, the vector field
$\vec{\tilde u}_0(\vec\theta)$ can be shown to satisfy the relation
\begin{equation}
  \vec{\tilde u}_0(\vec\theta) = \nabla \ln \bigl[ 1 -
  \kappa_0(\vec\theta) \bigr] = \nabla
  \tilde\kappa_0(\vec\theta) \; ,
\end{equation}
where $\kappa_0$ is the true dimensionless mass map and
$\tilde\kappa_0(\vec\theta) = \ln \bigl[ 1 - \kappa_0(\vec\theta)
\bigr]$. However, because of statistical and measurement errors,
$\vec{\tilde u}$ is not necessarily curl-free, and thus $\kappa$ can
be determined only approximately. In a separate paper
(Lombardi \& Bertin 1998b) we have shown that
\begin{itemize}
\item The statistical errors on $\kappa(\vec\theta)$ are minimized if
  this function is calculated as
  \begin{equation}
    \label{eq:H^SS}
    \tilde\kappa(\vec\theta) = \bar \kappa + \int_\Omega
    \vec H^\mathrm{SS}(\vec\theta, \vec\theta') \cdot \vec{\tilde
    u}(\vec\theta') \, \diff^2 \theta' \; ,
  \end{equation}
  where $\bar \kappa$ is a constant (introduced to take into account
  the \textit{sheet invariance\/}), $\vec H^\mathrm{SS}(\vec\theta,
  \vec\theta')$ is the noise-filtering kernel (Seitz \& Schneider
  1996), and $\Omega$ is the field of observation.
\item The same mass map can be obtained by solving the equations
  \begin{eqnarray}
    \label{eq:vN1}
    &&\nabla^2 \tilde\kappa = \nabla \cdot \vec{\tilde u} \; , \\
    \label{eq:vN2}
    &&\nabla \tilde\kappa \cdot \vec n = \vec{\tilde u} \cdot \vec n \qquad
    \mbox{on $\partial \Omega$} \; ,
  \end{eqnarray}
  where $\vec n$ is the unit vector perpendicular to the boundary
  $\partial \Omega$ of the field of observation $\Omega$. Hence, the kernel
  $\vec H^\mathrm{SS}(\vec\theta, \vec\theta')$ can be identified as
  the Green function of this Neumann boundary problem.
\item Equations \eqref{eq:vN1} and \eqref{eq:vN2} are precisely the
  Euler equations associated with the functional
  \begin{equation}
    \label{eq:S}
    S = \frac{1}{2} \int_\Omega \bigl\| \nabla
    \tilde\kappa(\vec\theta) - \vec{\tilde u}(\vec\theta) \bigr\|^2
    \diff^2 \theta \; .
  \end{equation}
  In other words, the functional $S$ is minimized when
  $\tilde\kappa(\vec\theta)$ is calculated using Eq.~\eqref{eq:H^SS}
  or, equivalently, by solving Eqs.~\eqref{eq:vN1} and \eqref{eq:vN2}.
\end{itemize}
To these three formulations of the mass inversion problem there
correspond three practical methods to calculate $\kappa(\vec\theta)$
from a given set of data.
\begin{itemize}
\item The first method considered is based on a direct calculation of
  the kernel $\vec H^\mathrm{SS}$ (Seitz \& Schneider 1996). Once
  this kernel has been calculated for a given field $\Omega$, the mass
  inversion is straightforward (note that the kernel $\vec
  H^\mathrm{SS}$ depends on the field of observation). A problem with
  this method is that a calculation of $\vec H^\mathrm{SS}$ is
  expensive in terms of memory requirements and computation time. In
  fact, in order to compute $\kappa$ on a square grid of $N \times N$
  points, $\vec H^\mathrm{SS}$ must be calculated on a
  multidimensional grid of $N \times N \times N \times N$ points.
  Moreover, $2N^4$ multiplications are needed to evaluate
  Eq.~\eqref{eq:H^SS}, and thus the method is of order $\O \bigl( N^4
  \bigr)$. Because of the large memory needed to allocate $\vec
  H^\mathrm{SS}$, calculations can be performed only with a limited
  value of $N$ (typically $N \sim 50$).
\item The introduction of a method that directly solves the Neumann
  problem allows one to go beyond many of the limitations of the $\vec
  H^\mathrm{SS}$ method. Equations~\eqref{eq:vN1} and \eqref{eq:vN2}
  can be solved using an \textit{over-relaxation method} (Seitz \&
  Schneider 1998). In this case $\tilde\kappa(\vec\theta)$ is
  calculated directly, and thus we need to allocate only $N \times N$
  real numbers. Moreover, the method can be applied without
  difficulties to ``strange'' geometries $\Omega$ (while the previous
  method is straightforward only when applied to rectangular or
  circular fields).  The over-relaxation method is quicker than the
  kernel method, being approximately of order $\O\bigl( N^3 \bigr)$.
\item A \textit{direct\/} method to minimize the functional
  \eqref{eq:S} will be presented in this paper. As we will see, this
  method has several advantages and turns out to be very efficient
  from a computational point of view, being of order $\O \bigl( N^2
  \log N \bigr)$. Moreover, it is extremely easy to implement (two
  implementations for rectangular fields $\Omega$ written in {\tt C}
  and in {\tt IDL} are freely available on request).
\end{itemize}

We should stress that, as proved in an earlier paper (Lombardi \&
Bertin 1998b), the three formulations are mathematically equivalent.
Thus it would not be surprising to find that proper numerical
implementations perform, for large values of the grid number $N$, in a
similar manner as far as accuracy and reliability are concerned. In
practice, for finite values of $N$, the third method turns out to be
characterized by small errors, often smaller than those associated
with the other two procedures.

\section{A direct method to solve the variational principle}

Direct methods in variational problems are well-known especially in
applied mathematics (see, e.g., Gelfand \& Fomin 1963). Suppose that
one can find a \textit{complete\/} set of functions $\{ f_\alpha \}$
on the domain $\Omega$ (the full definition of ``complete'' will be
given below), so that any function on $\Omega$ can be represented as a
linear combination of the form
\begin{equation}
  \label{eq:K}
  \tilde\kappa(\vec\theta) = \sum_{\alpha=1}^\infty c_\alpha
  f_\alpha(\vec\theta) \; . 
\end{equation}
More precisely, we assume that for any function $\tilde
\kappa(\vec\theta)$, there is a choice for the coefficients $\{
c_\alpha \}$ such that
\begin{equation}
  \label{eq:complete2}
  \int_\Omega \biggl\| \nabla \tilde\kappa(\vec\theta) -
  \sum_{\alpha=1}^\infty c_\alpha \nabla f_\alpha(\vec\theta) \biggr\|^2
  \, \diff^2 \theta = 0 \; .
\end{equation}
Let us now introduce a sequence of trial mass maps
\begin{equation}
  \label{eq:K^[n]}
  \tilde\kappa^{[n]}(\vec\theta) = \sum_{\alpha=1}^n c_\alpha^{[n]}
  f_\alpha(\vec\theta) \; .
\end{equation}
We further require that the function $\tilde\kappa^{[n]}$ minimizes the
functional $S$: in other words, $c^{[n]}_1, c^{[n]}_2, \dots,
c^{[n]}_n$ are chosen so that the functional $S$ has minimum value.
This obviously happens when
\begin{equation}
  \frac{\partial S}{\partial c^{[n]}_\alpha} = 0 \qquad \mbox{for
  $\alpha = 1, 2, \dots, n$} \; .
\end{equation}
Solving this set of $n$ equations, we obtain the $n$ coefficients
$c^{[n]}_\alpha$, and thus the function $\tilde\kappa^{[n]}$. By
repeating this operation for a sequence of values of $n$, we find a
sequence of functions $\tilde\kappa^{[n]}$. These functions, under
suitable assumptions (verified in our problem), have the following
properties (see Gelfand \& Fomin 1963 for a detailed discussion): (i)
Let us call $S^{[n]}$ the value of $S$ when $\tilde\kappa$ is replaced
by the function $\tilde\kappa^{[n]}$. Then, obviously, the sequence
$S^{[n]}$ is not increasing. (ii) If the set $\{ f_\alpha \}$ is
complete, then the functions $\tilde\kappa^{[n]}$ converge to the
solution $\tilde\kappa$ of the problem. This method thus provides a
way to obtain the function $\tilde\kappa(\vec\theta)$ with desired
accuracy.

The method described here can be easily applied to our problem. In
fact, by expanding $\tilde\kappa^{[n]}(\vec\theta)$ as in
Eq.~\eqref{eq:K^[n]}, we find
\begin{equation}
  \label{eq:system}
  \frac{\partial S}{\partial c^{[n]}_\alpha} = \int_\Omega
  \nabla f_\alpha(\vec\theta) \cdot \biggl[ \sum_{\beta=1}^n
  c^{[n]}_\beta \nabla f_\beta(\vec\theta) - \vec{\tilde u}(\vec\theta)
  \biggr] \diff^2 \theta = 0 \; .
\end{equation}
The previous equation, for $\alpha = 1, 2, \dots, n$, represents a
linear system of $n$ equations for the $n$ variables $\bigl\{
c^{[n]}_\alpha \bigr\}$. Its solution is thus the set of coefficients
to be used in Eq.~\eqref{eq:K^[n]}.

However, we note that care must be taken in the choice of a
\textit{complete\/} set of functions. Let us define, for the purpose,
the product $\langle \vec v, \vec w \rangle$ between two generic
vector fields $\vec v(\vec\theta)$ and $\vec w(\vec\theta)$ as
\begin{equation}
  \label{eq:<f,g>}
  \langle \vec v, \vec w \rangle = \int_\Omega \vec v(\vec\theta)
  \cdot \vec w(\vec\theta) \diff^2 \theta \; .
\end{equation}
As our problem involves $\nabla \tilde\kappa$, the completeness has to
be referred to the set of the gradients. In other words, the set
$\bigl\{ f_\alpha \bigr\}$ is complete if
\begin{equation}
  \label{eq:complete}
  \langle \nabla f_\alpha, \nabla \tilde \kappa \rangle = 0
\end{equation}
for every $\alpha$ implies $\tilde\kappa(\vec\theta) =
\mbox{constant}$. It is easy to show that this condition is equivalent
to Eq.~\eqref{eq:complete2}.

The direct method can be further simplified if a set of functions $\{
f_\alpha \}$ can be taken to satisfy a suitable orthonormality
condition, so that the gradients of the functions $\{ f_\alpha \}$
satisfy
\begin{equation}
  \label{eq:ortho}
  \langle \nabla f_\alpha, \nabla f_\beta \rangle =
  \delta_{\alpha\beta} \; ,
\end{equation}
where $\delta_{\alpha\beta} = 1$ for $\alpha = \beta$ and $0$
otherwise. Then Eq.~\eqref{eq:system} can be rewritten simply as
\begin{equation}
  c^{[n]}_\alpha = \langle \nabla f_\alpha, \vec{\tilde u} \rangle \;
  .
\end{equation}
Thus, with the use of an orthonormal set of functions we have secured
two important advantages: (i) The linear system \eqref{eq:system} has
been diagonalized, so that its solution is trivial. (ii) The
coefficients $c^{[n]}_\alpha$ no longer depend on $n$: that is, the
coefficients of the \textit{exact\/} solution are given by $c_\alpha =
\langle \nabla f_\alpha, \vec{\tilde u} \rangle$.

Because of these advantages, whenever possible an orthonormal set of
functions should be used. We note, however, that the orthonormality
condition \eqref{eq:ortho} depends on the field of observation
$\Omega$. Even if the existence of an orthonormal set of functions is
always guaranteed by the spectral theory for the Laplace operator (see
Brezis 1987), for ``strange'' geometries, it may be \textit{non
  trivial\/} to find a complete orthonormal set of functions. In such
cases, we need to solve the linear system \eqref{eq:system}.

The direct method described above has several advantages with respect
to the ``kernel'' method and to the over-relaxation method: (i) The
method is fast in the case where an orthonormal set of functions can
be found. In fact, we need only to evaluate one integral for each
coefficient $c_\alpha$ that we want to calculate.  (ii) The method
does not require a large amount of memory: we need to retain only the
$n$ values of the coefficients $c_\alpha$.  (iii) The precision of the
inversion is driven in a natural way by the value of $n$. Typically,
the larger $\alpha$ is, the smaller the length scale of $f_\alpha$
(see below). (iv) In some cases, the decomposition of the mass density
$\tilde\kappa(\vec\theta)$ in terms of the functions $f_\alpha$ can be
useful.

\section{Rectangular fields}

When the field $\Omega$ is rectangular, an orthonormal set of
functions can be written easily. Here we consider the special case
when $\Omega$ is a square of length $\pi$ (in some suitable units);
any rectangular field can be handled in a similar manner. In the case
considered, an orthonormal set of functions is given by
\begin{equation}
  \label{eq:f_ab}
  f_{\alpha\beta}(\vec\theta) = n_{\alpha\beta} \cos \alpha \theta_1
  \cos \beta \theta_2 \; ,
\end{equation}
with $(\alpha, \beta) \in \mathbb{N}^2 \setminus (0,0)$. The
normalization $n_{\alpha\beta}$ is defined as
\begin{equation}
  n_{\alpha\beta} = 
  \cases{
    \frac{\displaystyle{\sqrt{2}}}{\displaystyle{\pi \sqrt{\alpha^2 + 
        \beta^2}}} & for $\alpha = 0$ or $\beta = 0 \; ,$ \cr
    \frac{\displaystyle{2}}{\displaystyle{\pi \sqrt{\alpha^2 +
        \beta^2}}} & otherwise$\; .$ }
\end{equation}
The function $f_{00}$ is not defined. Note that here we use two
indices for the set. Cosines must be used in order to have a
\textit{complete\/} set (see Eqs.~\eqref{eq:complete2},
\eqref{eq:complete}, and Appendix~A). Our problem is solved in terms
of the coefficients $c_{\alpha\beta}$:
\begin{eqnarray}
  &c_{\alpha\beta} =& -n_{\alpha\beta} \int_\Omega \bigl[ \alpha
  \tilde u_1(\vec\theta) \sin \alpha \theta_1 \cos \beta \theta_2 +
  \nonumber\\
  && \phantom{-n_{\alpha\beta} \int_\Omega \bigl[ \,}
  \beta \tilde u_2(\vec\theta) \cos \alpha \theta_1 \sin \beta
  \theta_2 \bigr] \diff^2 \theta \; , 
  \label{eq:c_ab} \\
  &\tilde\kappa(\vec\theta) =& \sum_{\alpha\beta} c_{\alpha\beta}
  n_{\alpha\beta} \cos \alpha \theta_1 \cos \beta\theta_2 \; .
  \label{eq:kappa_ab}
\end{eqnarray}
We now observe that the particular choice of the orthonormal set
$\{ f_{\alpha\beta} \}$ allows us to use fast Fourier transform (FFT)
techniques to evaluate Eqs.~\eqref{eq:c_ab} and \eqref{eq:kappa_ab}.
The use of FFT makes the direct method very efficient: in particular
the method becomes of order $\O\bigl( N^2 \log N \bigr)$. Moreover,
several optimized FFT libraries are available.

The optimal truncation for the series \eqref{eq:kappa_ab} is
determined by the adopted grid numbers: for a grid of $N \times M$
points, $\alpha$ should run from $0$ to $N-1$, and $\beta$ from $0$ to
$M-1$ (this is standard practice for FFT libraries).

\section{Performance}

Our method has been implemented in {\tt C} and in {\tt IDL}. The {\tt
  C} version uses the library {\tt FFTW} (``Fastest Fourier Transform
in the West,'' version $2.0.1$) to perform discrete Fourier transforms
(DFT). This library, written by Matteo Frigo and Steven G. Johnson, is
considered the quickest DFT library publicly available.  The
performance of our direct method is compared with that of the
over-relaxation method, also implemented in {\tt C}. The procedure
used in the tests is summarized in the following points:
\begin{enumerate}
\item A simple model for the dimensionless mass distribution has been
  chosen. Then the mass distribution $\tilde\kappa_0$ is calculated on
  a grid of $N \times N$ points.
\item The associated field $\vec{\tilde u}_0$ is calculated on the
  same grid using a $3$-point Lagrangian interpolation in order to
  numerically evaluate the derivatives that are needed.
\item Noise is added to the vector field $\vec{\tilde u}_0$ using an
  analytical model for the noise derived earlier (Lombardi \& Bertin
  1998b). In practice, the various Fourier components of the noise are
  added using a suitable model for the power spectrum.
\item The resulting noisy $\vec{\tilde u}$ map is inverted using the
  over-relaxation method and the present direct method. The two
  dimensionless mass maps obtained are then compared. Moreover, the
  inversion times are recorded.
\end{enumerate}

The results obtained in the tests are the following:

\begin{itemize}
\item The two mass densities obtained are consistent with each other.
\item Because of the set of functions used, the errors produced by the
  direct method are larger on the boundary of the field. For this
  reason, we suggest that a one pixel strip around the field should be
  discarded. The area discarded is very small.
\item Some tests have been performed by providing $\vec{\tilde u}_0$
  to the inversion procedures. This allows us to compare the
  reconstructed mass density with the original map $\kappa_0$. From
  such tests we have noted that the discretization errors of the
  direct method are slightly smaller than the discretization errors of
  the over-relaxation method.
\item The results of the two methods differ because of the
  \textit{sheet invariance\/}: in particular, the direct method always
  gives a ``reduced'' mass map $\tilde\kappa$ with vanishing total
  mass.
\end{itemize}
Regarding the second item, we note that the error is related to the
finite sampling scale of the method; the error affects only the
outermost pixel because of the proper choice of the truncation (see
comment at the end of Sect.~3).

\begin{figure}[!t]
  \resizebox{\hsize}{!}{\includegraphics{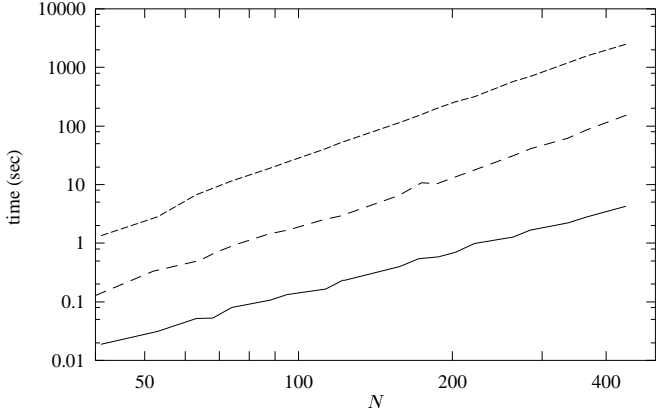}}
  \caption{Execution time per call vs.\ grid number $N$. The solid line
    refers to the direct method applied to ``good numbers'' (values
    $2N-1$ that can be factorized with small primes), the long-dashed
    refers to the direct method applied to ``bad numbers'' ($2N-1$
    prime), and the short-dashed line to the over-relaxation method.}
  \label{fig:1}
\end{figure}

The measured execution times are plotted in Fig.~1 for different
values of $N$. These are the averaged CPU execution times for a single
reconstruction on a SUN Ultra~1 workstation. From this figure it is
clear that the direct method is much faster than the over-relaxation
method. Here we should recall that, because of some characteristics of
the \texttt{FFTW} library, the execution time of the direct method can
change significantly even for neighbouring values of $N$. In
particular, the inversion is faster when $(2N - 1)$ can be factorized
with small prime numbers, and is slower in other cases (see Fig.~1).
For example, the execution time (on a SUN Ultra~1) changes from
$2.942$ to $0.232$ seconds when $N$ changes from $121$ to $122$.
Finally, we observe that our implementation of the direct method is
not optimal: in fact, with a different (non-trivial) use of FFT one
might gain an additional factor of 4 on the execution time.

Besides the appealing aspects of simplicity inherent to the direct
method described in this paper, we should note that gaining three
orders of magnitude in CPU time will make it possible to undertake a
few long-term projects of simulated observations (in particular, with
the goal of a statistically sound investigation of the quality of mass
reconstruction; but other objectives might be formulated, e.g. in the
cosmological context) that would remain practically out of reach for
other intrinsically slow reconstruction methods.

\section{Examples of simulated reconstructions}

\begin{figure}[!t]
  \resizebox{\hsize}{!}{\includegraphics{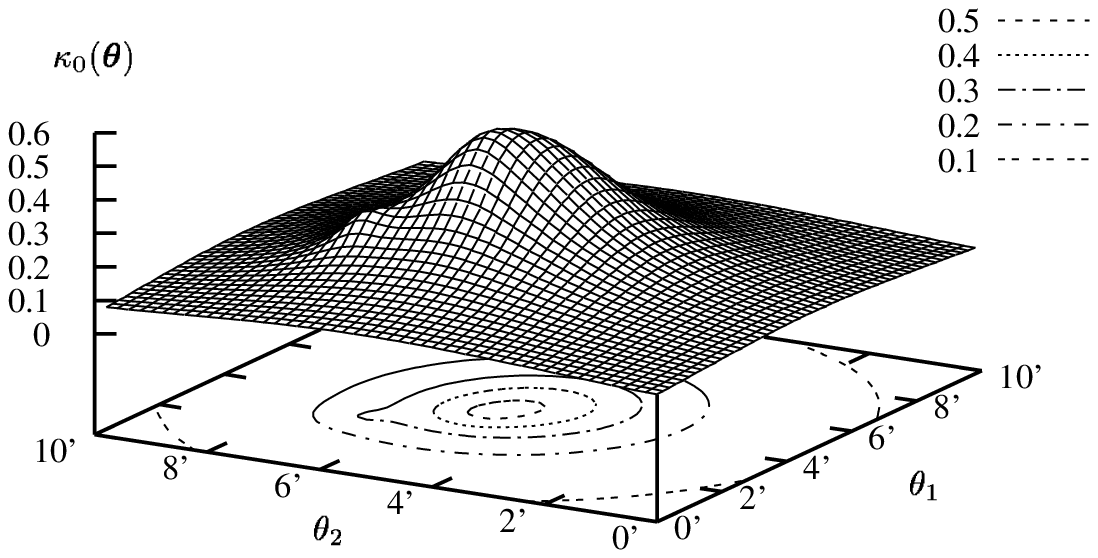}} 
  \resizebox{\hsize}{!}{\includegraphics{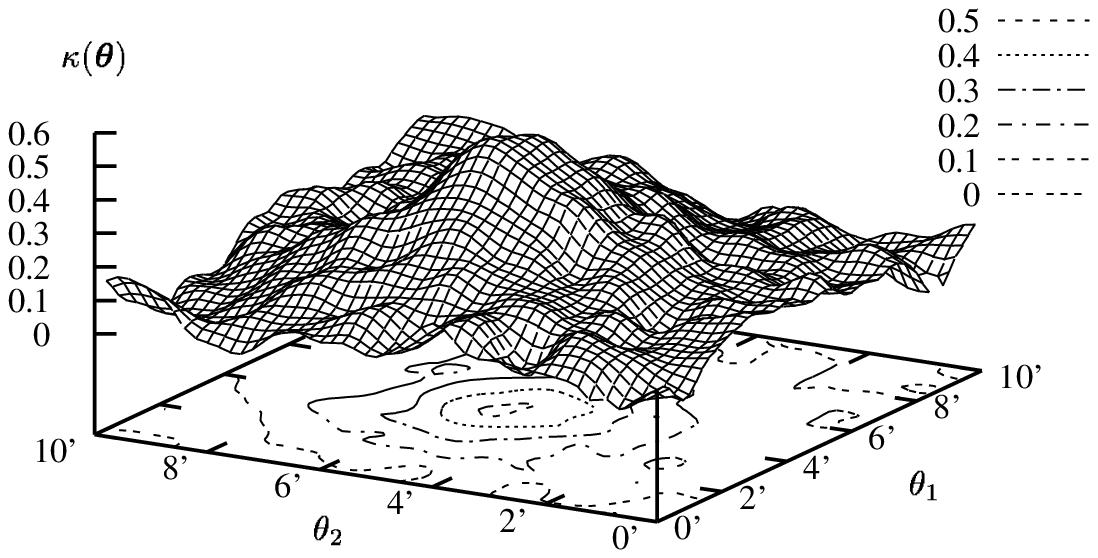}} 
  \resizebox{\hsize}{!}{\includegraphics{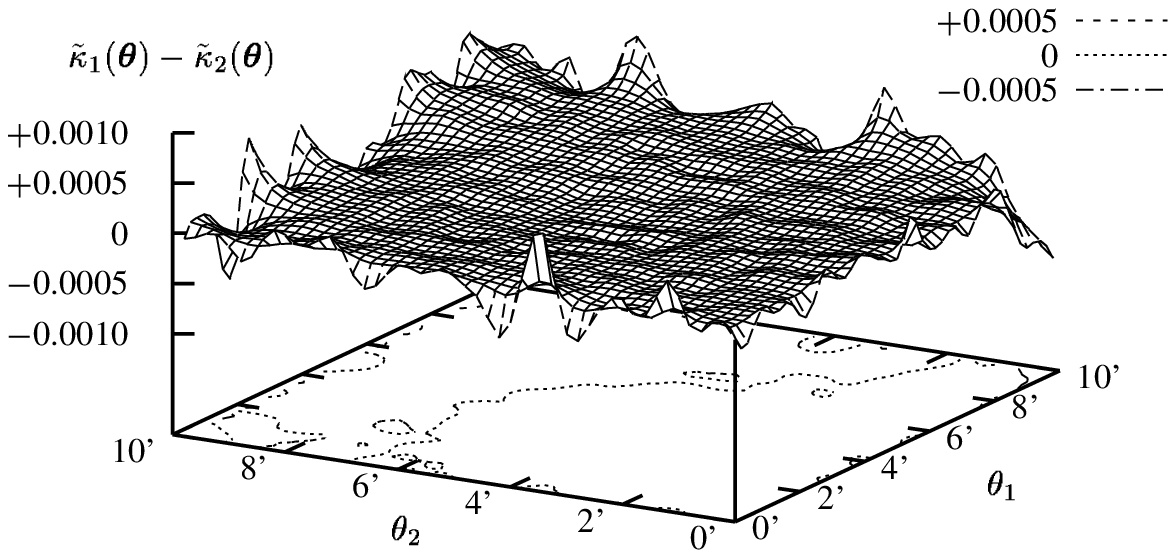}} 
  \caption{A typical result of mass reconstruction; at the adopted
    distance for the lensing cluster, the side of the square field, 10
    arcminutes, corresponds to approximately 2.88 Mpc. From top to
    bottom, true dimensionless mass distribution, reconstructed
    distribution (from the direct method), and difference between maps
    of the variable $\tilde\kappa$ derived from direct and
    over-relaxation methods.  The very small residuals show that the
    two methods are practically equivalent in terms of accuracy.}
  \label{fig:2}
\end{figure}

In addition to the reconstructions from ``synthetic'' data as
described in the previous Section, we have performed several
additional tests in order to demonstrate the reliability of our
method. The tests, designed with the aim to reproduce the main
features of a ``real'' reconstruction, have been made following a
straightforward procedure (see, e.g., Lombardi \& Bertin 1999 for
similar simulations).

First we have generated a population of source galaxies using a
pseudo-random number generator. Here a source galaxy is represented by
its position and by its ellipticity (see Eq.~(11) of Lombardi \&
Bertin 1999, following Seitz \& Schneider 1997). Positions are drawn
from a homogeneous distribution (with a density of $70$ galaxies per
square arcmin), while ellipticities are drawn from a truncated
Gaussian distribution with variance $\sigma^2 = (0.3)^2$.  Sources are
assumed to have all the same redshift $z_\mathrm{s} = 1.5$. Source
ellipticities are then transformed into observed ellipticities.  For
simplicity, the observed galaxy positions are assumed to be equal to
the source positions: in other words, no depletion effects are
included in the simulations.

Then the calculation of the observed ellipticities has been done by
referring to a cluster of galaxies placed at $z_\mathrm{d} = 0.3$ with
total mass inside the $10$'${} \times 10$' field $1.5 \times 10^{15}
\, M_\odot$. For the purpose of introducing the lensing effects, we
only need to specify the dimensionless projected mass map
$\kappa(\vec\theta)$.  For simplicity, we have used a density
distribution made of three symmetrical components; each component is
described by the analytical model outlined by Schneider \textit{et
  al}. (1992, p.~244), which, at large radii, is approximately
isothermal.

By averaging the observed ellipticities, we have then obtained a map
of the reduced shear $g(\vec\theta)$ and, from that map, the vector
field $\vec{\tilde u}(\vec\theta)$. The mass inversion has been
performed using the direct method and the over-relaxation method. The
two mass distributions have been then compared.

An example of typical results is shown in Fig.~2. Here, from top to
bottom, we display the original cluster mass distribution, the
reconstruction obtained using the direct method, and the residuals,
i.e.\ the difference between the reconstructed maps from the direct
method and from the over-relaxation method. As the figure clearly
shows, differences are mainly confined to the boundary of the field
where they are found to be of the order of $0.0002$, well below the
statistical errors of the reconstruction.  In the inner field the
differences are about one order of magnitude smaller. Note also that
the wavy overall appearance of the reconstructed map is normal for
weak lensing reconstructions, resulting from the relatively low number
of source galaxies involved (see Lombardi \& Bertin 1998b for a
discussion of the statistical aspects of the problem).

\begin{acknowledgements}
  We thank Luigi Ambrosio and Peter Schneider for interesting
  discussions and suggestions. The reconstruction code uses
  \texttt{FFTW 2.0.1} by Matteo Frigo and Steven G. Johnson.  This
  work has been partially supported by MURST and by ASI of Italy.
\end{acknowledgements}

\begin{appendix}

\section{Completeness of $\{ f_{\alpha\beta} \}$}

In this appendix we will verify explicitly that the set of functions
defined in Eq.~\eqref{eq:f_ab} is complete, in the sense that
Eqs.~\eqref{eq:vN1} and \eqref{eq:vN2} can be recovered. For the
purpose, we will apply the Fourier theorem (Brezis 1987). In the
following, we will assume that $\vec{\tilde u}(\vec\theta)$ is a
\textit{smooth\/} vector field (we stress that this condition is
needed only for the proof provided below; the method remains applicable
to more general cases).

Let us consider a solution of the form \eqref{eq:K}. Then, because of
the orthonormality condition \eqref{eq:ortho}, we have $\langle \nabla
f_\alpha, \nabla \tilde\kappa \rangle = c_\alpha = \langle \nabla
f_\alpha, \vec{\tilde u} \rangle$, so that
\begin{equation}
  \langle \nabla f_\alpha, \nabla \tilde\kappa - \vec{\tilde u}
  \rangle = \int_\Omega \nabla f_\alpha \cdot (\nabla \tilde \kappa -
  \vec{\tilde u}) \, \diff^2 \theta = 0 \; .
\end{equation}
Now we observe that the previous equation holds for any $\alpha$: then
it holds also for any linear combination $f = \sum_\alpha d_\alpha
f_\alpha$ of $\{ f_\alpha \}$. Thus
\begin{eqnarray}
  &0 =& \int_\Omega \nabla f \cdot (\nabla \tilde\kappa - \vec{\tilde
    u}) \diff^2 \theta = 
  -\int_\Omega f \nabla \cdot (\nabla \tilde\kappa - \vec{\tilde
    u}) \, \diff^2 \theta \nonumber\\
  && {} + \int_{\partial \Omega} f (\nabla \tilde\kappa - \vec{\tilde
    u}) \cdot \vec n \, \diff \theta \; .
  \label{eq:App}
\end{eqnarray}
In the last step we have integrated by parts ($\vec n$ is the unit
vector orthogonal to the boundary $\partial \Omega$ of $\Omega$). 

We now use this equation to show that the chosen set of functions,
described by Eq.~\eqref{eq:f_ab}, is \textit{complete}, while, e.g., a
similar set made of sine functions would not be complete. By the
nature of the chosen set of functions we already know that we can
properly represent any smooth function $f$. Using this property, we
want to show that the two terms $\nabla \cdot (\nabla \tilde\kappa -
\vec{\tilde u})$ and $(\nabla \tilde\kappa - \vec{\tilde u}) \cdot
\vec n$ entering in the r.h.s.\ of Eq.~\eqref{eq:App} vanish on
$\Omega$ and on $\partial\Omega$ respectively.

For the purpose, we observe that if cosines are used as set of
functions, we can ``build'' any function $f$ provided that the
function has periodic derivatives on the boundary. In particular, if
$A \subset \Omega$ is an arbitrary open subset of $\Omega$, there is a
function $f$ that is positive on $A$ and vanishes on $\Omega \setminus
A$. Now suppose \textit{per absurdum\/} that the solution obtained
from the direct method does not satisfy Eq.~\eqref{eq:vN1}, so that,
e.g., $\nabla \cdot (\nabla \tilde\kappa - \vec{\tilde u}) > 0$ on a
point $\vec\theta^* \in \Omega$. Then, for the sign persistence
theorem, this quantity must be strictly positive in a neighborhood $A$
of $\vec\theta^*$. However, if we take a function $f$ which is
positive on $A$ and vanishes elsewhere, the rhs of Eq.~\eqref{eq:App}
will be positive, while the lhs vanishes, which is contradictory.
This proves that Eq.~\eqref{eq:vN1} is verified by cosines.

In a similar manner, now that we have ``disposed of'' the first term,
we observe that using cosines we can build a function $f$ that
vanishes everywhere on the boundary of $\partial \Omega$ except for a
neighborhood. In other words, given an open subset $B \subset \partial
\Omega$ of the boundary $\partial \Omega$, there is a function $f$
that is positive on $B$ and vanishes on $\partial \Omega \setminus B$.
Note that this property would not be satisfied if the set of functions
$\{ f_\alpha \}$ were based on sines. Using a proof similar to the one
given above, we obtain that $(\nabla \tilde\kappa - \vec{\tilde u})
\cdot \vec n$ must vanish, thus leading to Eq.~\eqref{eq:vN2}.

One might worry that, on the boundary, the chosen set of functions of
Eq.~\eqref{eq:f_ab} has zero derivative in the direction of $\vec n$,
i.e.\ $\nabla f_{\alpha\beta} \cdot \vec n = 0$. This might suggest
that the boundary condition \eqref{eq:vN2} cannot be reproduced. In
reality, although this is true \textit{pointwise}, this does not
affect the convergence in $\mathcal{L}^2$ that is relevant for our
problem (see Eqs.~\eqref{eq:complete2} and \eqref{eq:complete}). This
important point is best illustrated by the following example.

Suppose that we measure a constant field $\vec{\tilde u}(\vec\theta) =
(1, 0)$ on a square field $\Omega$ of side $\pi$. The obvious solution
for $\tilde\kappa$ in this case is $\tilde\kappa(\theta\vec) =
\theta_1$. Suppose now that we try to use a set of functions made of
sines. Then the corresponding coefficients $c_{\alpha\beta}$ would be
proportional to the integrals $\alpha \int_0^\pi \theta_1 \cos
\alpha\theta_1 \, \diff \theta_1 \int_0^\pi \sin \beta\theta_2 \,
\diff \theta_2 = 0$. Hence, every coefficient $c_{\alpha\beta}$
vanishes. This proves that a set based on sines is not complete
(condition ~\eqref{eq:complete} is not satisfied or, equivalently, a
curl-free vector field $\vec{\tilde u}$ leads to a vanishing mass
map). On the other hand, if we use the set~\eqref{eq:f_ab}, the
coefficients $c_{\alpha\beta}$ do not vanish and the corresponding
mass distribution is given by
\begin{equation}
  \tilde\kappa(\vec\theta) = - \sum_{\alpha \mbox{\scriptsize\ odd}}
  \frac{4}{\pi\alpha^2} \cos \alpha\theta_1 \; . 
\end{equation}
This function can be shown to reduce to $\tilde\kappa(\vec\theta) =
\theta_1 - \pi/2$.

\end{appendix}

\end{document}